\documentclass[sigconf,table,dvipsnames,table,10pt]{acmart} %

\usepackage[utf8]{inputenc}
\usepackage[T1]{fontenc}
\usepackage{fontawesome}
\usepackage{booktabs}

\usepackage{amsmath}
\usepackage{multirow}
\usepackage{makecell}
\usepackage{threeparttable}
\usepackage{tablefootnote}
\usepackage{graphicx}
\usepackage{setspace}%
\usepackage[margin=8pt,skip=5pt,belowskip=0pt,font={small,stretch=0.9},labelfont=bf]{caption}%
\usepackage{subcaption}

\usepackage{pgf-umlsd}

\usepackage{xspace}
\usepackage{xcolor}
\usepackage{pifont}
\usepackage[abbreviations]{foreign}  %
\usepackage[binary-units,per-mode=symbol]{siunitx}
\usepackage{ifthen}
\usepackage{tcolorbox}

\makeatletter
\DeclareRobustCommand*\cal{\@fontswitch\relax\mathcal}
\makeatother

\newboolean{showcomments}
\setboolean{showcomments}{true}
\ifthenelse{\boolean{showcomments}}
{ \newcommand{\mynote}[3]{
   \fbox{\bfseries\sffamily\scriptsize#1}
   {\small$\blacktriangleright$\textsf{\emph{\color{#3}{#2}}}$\blacktriangleleft$}}}
{ \newcommand{\mynote}[3]{}}

\newcommand{\sys}{\textsc{Hermes}\xspace}
\graphicspath{{./}{figures/}}

\usepackage{cleveref}

\usepackage{censor}

\copyrightyear{2020}
\acmYear{2020}
\acmConference[DEBS '20]{DEBS '20: 14th ACM International Conference on Distributed and Event-Based Systems}{July 13--17, 2020}{Montreal, Quebec, Canada}
\acmBooktitle{DEBS '20: 14th ACM International Conference on Distributed and Event-Based Systems, July 13--17, 2020, Montreal, Quebec, Canada}

\setcopyright{none}
\renewcommand\footnotetextcopyrightpermission[1]{}

\title{Hermes: Enabling Energy-efficient IoT Networks\\ with Generalized Deduplication}

\author{Christian G\"{o}ttel}
\affiliation{Department of Computer Science \\
University of Neuch\^atel, Switzerland}
\email{christian.goettel@unine.ch}

\author{Lars Nielsen}
\affiliation{DIGIT and Department of Engineering \\
Aarhus University, Denmark}
\email{lani@eng.au.dk}

\author{Niloofar Yazdani}
\affiliation{DIGIT and Department of Engineering \\
Aarhus University, Denmark}
\email{n.yazdani@eng.au.dk}

\author{Pascal Felber}
\affiliation{Department of Computer Science \\
University of Neuch\^atel, Switzerland}
\email{pascal.felber@unine.ch}

\author{Daniel E. Lucani}
\affiliation{DIGIT and Department of Engineering\\
Aarhus University, Denmark}
\email{daniel.lucani@eng.au.dk}

\author{Valerio Schiavoni}
\affiliation{Department of Computer Science \\
University of Neuch\^atel, Switzerland}
\email{valerio.schiavoni@unine.ch}

\begin{document}

\begin{abstract}
With the advent of the Internet of Things (IoT), the ever growing number of connected devices observed in recent years and foreseen for the next decade suggests that more and more data will have to be transmitted over a network, before being processed and stored in data centers.
Generalized deduplication (GD) is a novel technique to effectively reduce the data storage cost by identifying similar data chunks, and able to gradually reduce the pressure from the network infrastructure by limiting the data that needs to be transmitted.

This paper presents \sys, an application-level protocol for the data-plane that can operate over generalized deduplication, as well as over classic deduplication.
\sys significantly reduces the data transmission traffic while effectively decreasing the energy footprint, a relevant matter to consider in the context of IoT deployments. %
We fully implemented \sys and evaluated its performance using consumer-grade IoT devices (\eg, Raspberry Pi 4B models). 
Our results highlight several trade-offs that must be taken into account when considering real-world workloads.

\end{abstract}

 \keywords{IoT, generalized deduplication, energy efficiency}
\fancyhead{
     \vspace{-30pt}
     \begin{tikzpicture}
         \node[align=center] () at (0,0) {
             \begin{tcolorbox}[colback=yellow!40,
                               colframe=white,
                               width=\textwidth,
                               boxrule=0mm,
                               sharp corners]
	             \small
                     \centering
                     CC-BY 4.0. This is the author's preprint version of the camera-ready article. A shorter version of this paper is published in the proceedings of 14th ACM International Conference on Distributed and Event-Based Systems (DEBS 2020), \url{https://doi.org/10.1145/3401025.3404098}.
             \end{tcolorbox}
         };
     \end{tikzpicture}
}

\maketitle
\thispagestyle{fancy}

\section{Introduction}\label{sec:introduction}
The increasing adoption and expansion of Internet of Things (IoT) technologies is leading to an correspondingly growing number of connected, low-energy yet efficient and powerful Internet-enabled devices.
Predictions~\cite{idc-prediction} indicate \SI{175}{\zetta\byte} of data being produced by IoT devices already by 2025, with up to 1.25 Billion units deployed by 2030~\cite{columbus20182018} and 38\% of the global IP-based traffic generated by mobile devices~\cite{cisco2018cisco}.
Despite the imminent introduction of wider-band wireless technologies (\eg, 5G and beyond), it is clear that the pressure on the network will continue to increase.

Data compression~\cite{jarwan2019data}, deduplication~\cite{meyer2012study} or  network coding (NC) techniques~\cite{Rec2019} have been proposed to solve these IoT problems.
The latter is particularly interesting given the unreliable nature of data streams commonly found in real-world IoT deployments~\cite{amirjavid2016network}.
NC introduces redundancy where needed to protect against data loss, \ie, efficient protection of data.
On the other hand, compression and deduplication are interesting given the compression potential of IoT-generated data (\eg, smart power meters~\cite{morello2017smart}, weather stations~\cite{elijah2018overview}, bio-medical body sensors~\cite{catherwood2018community}).

\begin{figure*}[!t]
	\centering
	\includegraphics[width=0.75\linewidth]{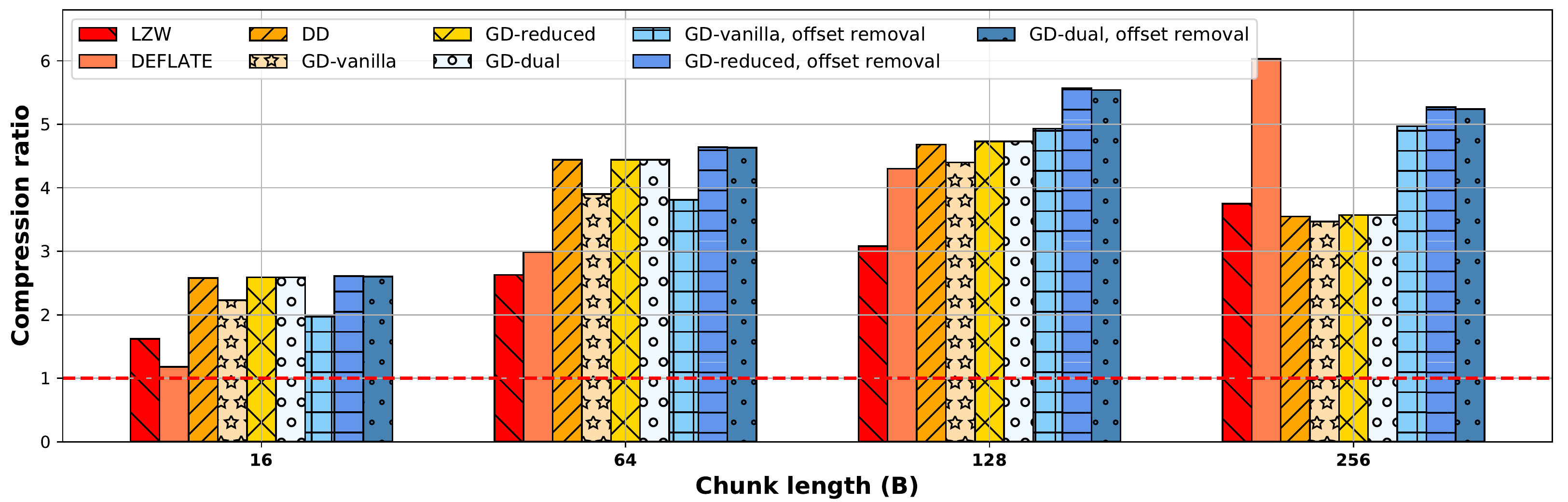}
	\caption{Compression ratio for the real world data set, water meter measurements, under different algorithms.}
	\label{fig:compression}
\end{figure*}

To validate this hypotesis, we applied different compression algorithms to a real-world dataset from the domain of ambient water and energy~\cite{water,batra2013s}.
\autoref{fig:compression} shows that there is a high potential to reduce data transmission (original size divided by compressed size - higher is better). 

The main challenge of standard compression algorithms in IoT is related to their processing costs~\cite{RFC1951-DEFLATE,7zip}.
Efficient compressors are usually too computationally intensive and memory-eager for IoT devices, while lightweight, memory-efficient approaches tend to have poorer compression performance~\cite{1055714,1055934,1659158}.

An added challenge is that many IoT applications rely on \emph{small data packets} and \emph{compress data on a per packet basis} due to memory limitations, which curbs the compression potential of standard compressors~\cite{yazdani2019protocols}. 
In fact, \autoref{fig:compression} shows that the compression ratio for two standard compression algorithms LZW~\cite{welch1984technique} and DEFLATE~\cite{RFC1951-DEFLATE} decreases dramatically for smaller chunk lengths. 

Network deduplication~\cite{spring2000protocol} is a well-known technique to reduce network traffic.
It operates by replacing a repeating byte sequence with a shorter hash value, which is later used to identify the intended content by the receiving side.
Sanadhya \etal~\cite{sanadhya2012asymmetric} proposed \emph{asymmetric caching}, where a source node performs deduplication on the outgoing chunks of data based on its cache content and the sink node's feedback, similar to what was originally done on Web caches in~\cite{spring2000protocol}.
A sink node sends timely feedback to the source node containing selected portion of its cache that is most likely to be useful to increase the probability of matching.
For instance, \cite{zhou2013efficient} describes a deduplication-based file communication system that leverage manifest feedback.
The source node splits each file into chunks and associated hash values.
Then, it checks locally for duplication based on its cache.
In case of misses, the hash values of missing duplicates are sent to the sink node for further duplication detection. 
The feedback packets from the sink node include the query information and the manifests of the chunks that have been hit at the sink node.
The manifests are the hash values, addresses and sizes of the chunks.
In~\cite{hua2015redundancy} a traffic deduplication approach is proposed to merge independent streams of the same video content on the Internet using a novel overlay network.

An enabled router can then merge and assign an identifier to each video.
In case of a match of video identifiers, the router will handle the merge.
The described approaches are typically designed for point-to-point transmission scenarios and using large data chunks, large hashes, local caching, and operating on files with data known \emph{a priori}.

For IoT applications generating smaller data chunks on the fly (\ie, the nature of data is not known beforehand) and limited memory/computation, existing state-of-the-art approaches are not suitable to deliver energy- and memory-efficient protocols.
Furthermore, a large source of compression potential in IoT comes from the massive amount of data sources compared to a standard approach, which needs to be considered in the system's design.

Another limitation of the state-of-the-art is that compression is provided by finding only equal data chunks.
If two chunks differ in even one bit value, they will be considered two different chunks.

Although techniques such as Rabin fingerprinting~\cite{rabin1981fingerprinting} can be used to split data in non-uniform chunks (\ie, different sizes) to detect similarities, these require significant added computation and memory to hold the computed similarity hashes.
This makes the approach unsuitable for the IoT.

Generalized deduplication~(GD)~\cite{vestergaard2019generalized} (further detailed in \S\ref{sec:background}) is a recently introduced scheme reduce the cost of storage not only by finding equal data chunks, but also by finding similar data chunks.
As in other lossless compression schemes, similarities between chunks are identified \emph{without} the need to carry out delta compression to a pool of previous chunks and \emph{without} relying on similarity hashes for dynamic chunking.
The latter would be impractical for the small amount of generated data in IoT devices.
In~\cite{yazdani2019protocols}, a lossless, multi-source data transmission compression approach inspired by the concept of GD was proposed to reduce the amount of data transmission.
\autoref{fig:compression} shows the compression ratio for data deduplication~(DD) and GD, showing that GD has the potential to outperform DD in IoT scenarios, but also that GD outperform LZW and DEFLATE for small packet sizes.

This paper introduces \sys, a protocol and a corresponding complete implementation for data transmission reduction in sensor networks, especially suited for resource-limited data nodes.
Its design principles are inspired by the schemes proposed in~\cite{yazdani2019protocols}, but also expanding this approach as well as making judicious adaptations to tackle core implementation and system aspects.
\sys allows multiple sources to share a common (and growing) data pool at the sink node, typically a Cloud- or Edge-based device. %
All source transmissions contribute to growing the knowledge pool.
Thus, spatial data correlations across multiple sources (\eg, similar temperature data at the same time across the same city, similar smart metering consumption of several households) as well as temporal correlations across multiple sources (\eg, same electricity readings of house A today as house B a year ago) can be exploited for reducing data transmission and allow for better compression at the sink node.
Each device then benefits from contributions from other devices in order to reduce their traffic.
\sys performs this without direct interactions between devices.
We implemented and experimentally evaluated \sys' performance with micro- and macro-benchmarks on Raspberry Pi 4B. %
In the best case scenario, we show reductions of 3 orders of magnitude of GD over DD, but also over per packet compression using LZW and DEFLATE. 

The rest of the paper is organized as follows.
In \S\ref{sec:background}, we introduce a theoretical background on generalized deduplication.
In \S\ref{sec:comm}, we study different communication mechanisms based on GD.
The \sys system and protocol is detailed in \S\ref{sec:hermes}. %
The experimental setup and the results of our experimental evaluation are presented in \S\ref{sec:evaluation}. 
We survey related work in \S\ref{sec:relwork}.
Finally, Section~\ref{sec:conclusion} concludes and presents future directions.

\section{Background}
\label{sec:background}

We begin by giving basic definitions and notations that are used throughout the rest of paper.

\subsection{Fingerprints and data deduplication}
\label{sec:DD}

We define $f(.)$  a function that associates a (nearly) unique fingerprint to its input, either using standard hash functions, \eg, SHA-1, SHA-256, or checksums, \eg, CRC32, MD5.
A deployment of \sys should settle on a specific $f(.)$  based on the need of low collision probability, \ie, for two different inputs to match the same fingerprint, but also the size of the fingerprint in relation to the amount of data transmitted per packet.
We note as well that the resulting compression gains are partially related (and limited) by the latter.
For example, a payload of $40$ bytes using a SHA-1 fingerprint of $20$ bytes will not compress beyond a factor of $2$.
Our evaluation shows the trade-offs for different $f(.)$ options.

Data deduplication~(DD) eliminates redundant data, removing copies of repeating data chunks.

Classic DD divides each piece of data into multiple data chunks, $C_i$, and stores each unique data chunk only once, by distinguishing between repeating data patterns and saving those only once.
A fingerprint is linked to each chunk.
Each file or piece of data would then be represented as a sequence of fingerprints for each of its chunks.
To recover the original data, one only needs to search for the chunk associated to each fingerprint and concatenate the data in the right order.

\subsection{Generalized deduplication}
\label{sec:GD}

Generalized deduplication~(GD)~\cite{vestergaard2019generalized} is a lossless data compression approach.
It operates by eliminating equal as well as similar data chunks. %
This is achieved \emph{without} comparing directly to previous chunks, but rather using a transformation function to systematically cluster similar data.
GD splits each piece of data into a series of equal-sized smaller chunks $C_i$'s and maps each chunk, $C_i$, onto a pair of basis, $b_i$, and associated deviation, $d_i$, by applying a transformation function.
For the transformation function, an error-correcting code~(ECC) can be used.
Each basis $b_i$ , which is larger than $d_i$", is assigned a fingerprint, $f(b_i)$.
The basis is saved only once.
For simplicity, we use the notation $f_{b_i}$ instead of $f(b_i)$. %
Rather than saving $C_i$, GD stores a pair $f_{b_i}$ and $d_i$.
Note that DD can be considered as a special case of GD where there is no deviation, $d_i = 0$, and $b_i=C_i$.

\emph{GD Example}.
Consider a chunk as having a shape and a color.
In this case, DD would provide a fingerprint for each chunk A with color $c_A$ and shape $s_A$.
GD can define a transformation that would split the chunk into the pair $(s_A,c_A)$ and proceed to deduplicate based on $s_A$ only.
That is, GD deduplicates all fragments that have the same shape $s_A$, since shape is bigger and requires more bits to be represented.
Each shape (our basis) will have a unique fingerprint.
The color information is simpler and requires few bits to represent it.
We will keep the color $c_A$ in the description of the data, next to the fingerprint pointing to the description of the shape.
Recovering the chunk involves fetching details about the shape (basis) and then apply the correct color (deviation).
Naturally, this increases the chances of mapping data with similar information (same shape).
In general, these are simply bit sequences and the matching potential depends on the transformation that splits into basis-deviation pairs.

\subsection{GD for efficient data transmission}
\label{sec:GDTrans}

GD can also reduce data transmission in a lossless manner~\cite{yazdani2019protocols}.
The idea is to apply GD at the source node to send the basis only if not available at the sink node (\eg, Cloud, Edge device).
At the source node, \eg, a sensor node, each chunk of data $C_i$ is then mapped onto a pair $(b_i, d_i)$. %
To reduce network overhead, the source node first transmits the associated basis fingerprint, $f_{b_i}$, \eg, a hash common to all the source nodes, and the deviation $d_i$.
The sink node checks whether it has the basis for the basis fingerprint or not.
If the basis for the basis fingerprint is already available at the sink node, it saves the data and sends back an acknowledgement.
At this point, the source node erases the associated basis and deviation from its memory.
Otherwise, the sink node sends a basis request and the source node sends the basis itself.
When receiving the acknowledgement from the sink, the source erases the basis from its memory.

This process can be generalized to transmit the information about more chunks in a single packet. %
Notice that all source nodes leverage the same hash function, each of them exploits all basis fingerprints available at the sink node, whether they were generated by the same source node or another source node.

We study possible communication mechanisms based on DD and GD in \autoref{sec:comm}.

\subsection{Transformation functions}

A variety of functions exist to create the mapping from $C_i$ to $(b_i, d_i)$.
We rely on error-correcting codes (ECC): $C_i$ is the \emph{codeword} and, by applying the decoding function of the ECC, the received \emph{message} is the basis.
The deviation $d_i$ carries the information about the difference between the codeword and the \emph{error-free codeword}.
The latter is created by encoding the basis $b_i$ using the ECC encoding function.

\textbf{Hamming codes.}
Hamming codes~\cite{moreira2006essentials} are a valid family of linear ECCs that can be used for the considered mapping~\cite{nielsen2019alexandria}.
Let $m$ be the number of parity bits, then the codeword and the message are of length $n = 2^m - 1$~bits and $k=2^m - m - 1$~bits, respectively.
We also define $n_B = \lceil \frac{n}{8} \rceil$ and $k_B = \lceil \frac{k}{8}\rceil$ as the byte-length of the codeword and message, respectively. 

Hamming codes can correct one bit errors.
In our context, this means that codewords are at most one bit away from the error-free codewords.
That is, we will systematically match chunks to others that have one bit difference, without comparing to previously received data chunks.
The location of the bit is specified in a \emph{syndrome vector} of length $m$ bits. 
Thus, $m$~bits are enough to represent the deviation for Hamming codes.
Since Hamming is a binary code, where a non-zero value must be a $\texttt{1}$, there is no need to save the content of the one bit error. 
By applying Hamming codes as the transformation function, all chunks mapped to a given basis are at most one bit away from the error-free chunk. 

Note that $n = 2^m - 1$, which would not use at least the last bit in the last byte.
Given the specific structure of Hamming codes, we consider chunks of length $n + 1$~bits to represent data received in bytes.
The Hamming transformation is performed in the first $n$ bits.
The remaining bit is left untouched as part of the deviation (concatenated with the $m$ deviation bits).
Thus, the resulting deviation for chunks of size $n+1$ bits would be $m+1$~bits in total.
During recovery, the steps is undone and the additional bit is appended to the $n$ bits.

\textbf{Reed-Solomon codes.}
Reed-Solomon codes~\cite{wicker1999reed} are valid ECCs to use for creating the mapping~\cite{vestergaard2019generalized}.
Reed-Solomon codes operate on a block of data treated as a series of symbols from a finite field of size $q$, $\mathbb{F}_q$.
A $RS_q(n_B,k_B)$ is a Reed-Solomon code where $q$ specifies the finite field that symbols are from, $\mathbb{F}_q$, and $n_B$ and $k_B$ are the symbol-size of the codeword and message, respectively, where $n_B = q - 1$ and $k_B< n_B$.
$t = \lfloor \frac{n_B-k_B}{2} \rfloor$ is the error correction capability of the code.
For $q=256$, symbols are $8$~bits in length (a byte).
Using a short version of the code~\cite{moreira2006essentials}, $n_B$ is more flexible, \ie, $k_B<n_B<q$.

The deviation can be computed by a bitwise XOR of the original chunk and the error-free chunk.
Then, the deviation is the location and the content of the non-zero symbols of the resulting sequence of the XOR.
Using these codes, it is impossible to predict a deviation as in Hamming codes.
It is however possible to  specify the maximum length of the intended deviation.
To achieve this, we consider the \emph{covering radius} metric of the code, $R(\mathcal{C})$.
$R(\mathcal{C})$ is the largest Hamming distance that any chunk might be from the associated error-free codeword.
Covering radius of a few Reed-Solomon codes are shown in \autoref{table:RC}.

Thus, we will need a maximum number of bits for representing the location and content of non-zero symbols given by $R(\mathcal{C}).\lceil \log_2 n_B \rceil$~bits and $R(\mathcal{C}).\lceil \log_2 q \rceil$~bits, respectively. 

\begin{table}[t!]
\centering
\caption{Covering radius of few Reed-Solomon Codes.}
\rowcolors{1}{gray!10}{gray!0}
\begin{tabular}{l|c}
\hline
\rowcolor{gray!45}
\textbf{Code} & \textbf{$R(\mathcal{C})$}\\
\hline

$RS_{256}(16,14)$ & 2\\

$RS_{256}(255,253)$ & 2\\

$RS_{256}(255,247)$ & 8\\

$RS_{256}(64,56)$ & 11\\
\hline
\end{tabular}
\label{table:RC}
\end{table}

\subsection{Preprocessing}

After splitting each piece of data into a series of equal-sized (and smaller) chunks $C_i$'s, but before applying a transformation function for mapping, we can apply an additional step to enhance the compression.
In this context, we think specifically on time-series and/or sensor data that contains a number of samples.
These samples could be from different sensors within the device (\eg, speed, vibrations, temperature) and/or for the same sensor over time.

\textbf{Delta encoding within a data packet.}
Considering each chunk $C_i$ as a concatenation of samples, applying delta encoding keeps the first sample unchanged and following samples will be replaced by the difference between the current sample and the previous one. 

Note that we consider delta encoding to be performed within each data packet and not across data packets.
This can reduce the range of the data (variance) if samples have constant or small variation.
The reduction in the data range can increase the matching probability of the bases.
On the other hand, delta encoding can increase by one bit (the sign) the required number of bits to represent the data samples. %
For example, a 3-byte sample $128,127,128$ would encode to $128,-1,1$.

\textbf{Offset removal within a data packet.}
To reduce the range of data and to keep the number of bits per sample unchanged, we propose the idea of \emph{offset removal}.
Let us consider each chunk $C_i$ as a concatenation of samples. 
We determine the minimum sample value of each chunk $C_i$.
Then, by applying offset removal, each sample can be represented as the differential value between the current sample and the minimum value of the chunk.
We save the minimum value as part of the associated deviation (\ie, we increase the deviation by an additional value).
The 3-byte sample $128,127,128$  would then encode to $1,0,1$ with a minimum value of $127$ added to the deviation.

\begin{figure}[!t]
  \centering
  \includegraphics[width=0.8\linewidth]{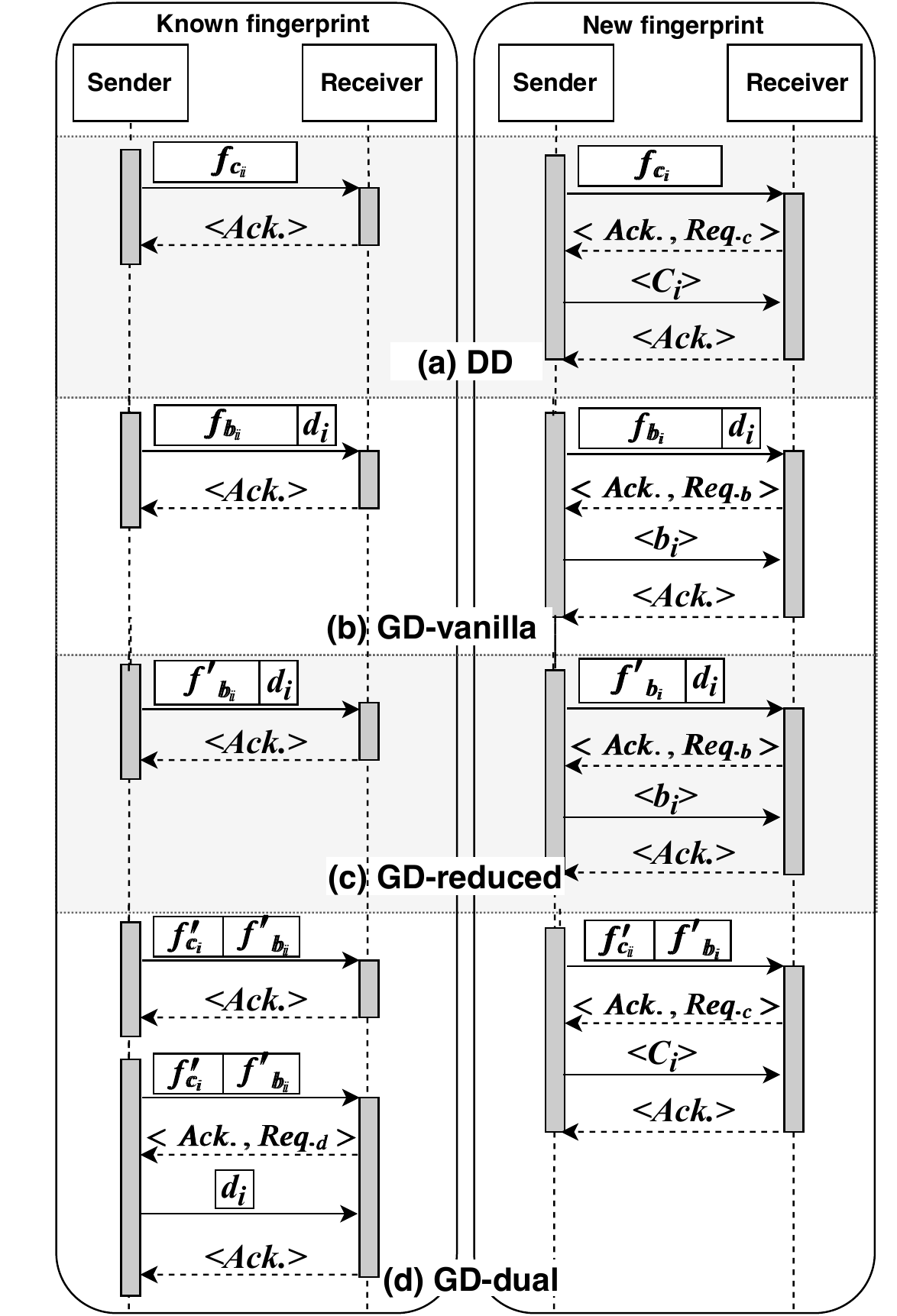}
  \caption{Communication mechanisms.}
  \label{fig:transmission}
\end{figure}

\section{Communication Mechanisms}
\label{sec:comm}

We devise three different communication mechanisms on top of GD, depicted in \autoref{fig:transmission}.

For each of them we describe the trade-offs and advantages with respect to DD.

\textbf{Baseline deduplication (DD):}
\autoref{fig:transmission}~(a) presents the baseline mechanism on top of DD .

Each fingerprint corresponds to a unique chunk value (with high probability given the fingerprint function).

\textbf{Baseline gen. deduplication (GD-vanilla):}

A similar mechanism can be operated on top of GD, as shown in \autoref{fig:transmission}~(b).
Notice that DD and GD-vanilla use the same fingerprint length.
Thus, GD-vanilla may incur a slightly larger overhead than DD if an exact duplicate chunk is already in the sink node, since the transmission of the deviation would be redundant.

\textbf{Reduced fingerprint gen. deduplication (GD-reduced):}
\autoref{fig:transmission}~(c) shows a variant that compensates the overhead of GD-vanilla . %
Since the bases for GD have fewer bits than the chunk length, we can consider a slightly smaller fingerprint for GD. %
In GD-reduced, we design the system such that the length of the fingerprint plus the deviation is equal to the length of the fingerprint for DD.
This removes the penalty when transmitting exact duplicates, albeit potentially compromising the probability of collision of the fingerprint, \ie, having a larger probability of two different bases being mapped to the same fingerprint.

\textbf{Dual fingerprint generalised deduplication (GD-dual):}
Finally, we describe a hybrid approach that allows the sink to identify whether it sees an exact replica or a new chunk potentially associated to a previously seen basis.
As shown in \autoref{fig:transmission}~(d), this approach transmits the fingerprint of the chunk, $f_{c_i}^\prime$, and of the basis, $f_{b_i}^\prime$, at the same time.
We set each of these as half length of $f_{c_i}$ of the classic DD approach.
If the chunk is already available in the sink node~(receiver), the sink sends an acknowledgement.
Otherwise, it checks if the basis has been received.
If it is already available, the sink node sends a request for deviation.
Else, the sink sends a request for the chunk.
After receiving the deviation or the chunk, the sink node sends back an acknowledgement.
Notice that the probability of collision is equal for both communication mechanisms DD and GD-dual.
This is due to the fact that the total fingerprint length in bits of DD is the same as the total length the two fingerprints (\eg, calculated with the same algorithm and only sending a fraction of the bits).
Notice that if a given chunk is already available at the sink node, both $f_{b_i}^\prime$ and $f_{c_i}^\prime$ should match.
On the other hand, if a similar chunk matches to the same basis, $f_{b_i}^\prime$ will match, but not $f_{c_i}^\prime$.
However, it is possible to calculate the chunk's fingerprint locally, due to availability of both basis and deviation to make sure it matches the received $f_{c_i}^\prime$.
Thus, GD-dual provides a lower probability of collision than GD-reduced and equivalent to DD and GD-vanilla.

\subsection{Transmission cost}
\label{sec:tranCostAna}

DD and GDD schemes have different transmission (\ie, network) costs.

The parameter $C$ denotes the total number of chunks transmitted by the source.
Let us define ${\cal B}_{GD}$ and ${\cal B}_{DD}$ as the total number of different bases generated by a source node out of $C$ data chunks by using GD and DD, respectively, where a ``basis'' for DD is equal to the chunk itself (\S\ref{sec:DD}).
We also define $d_B(i)$ and $h_{B,\cal P}$ as the deviation byte length of $C_i$ and a basis fingerprint's byte length for the scheme ${\cal P}$, respectively.
Accordingly, the transmission cost $T_{{\cal P}} (C)$ for a source node depends on the scheme ${\cal P}$ where, 
\vspace{-5pt}
\begin{equation}
\label{eq:DDcost}
T_{DD}(C) = {\cal B}_{DD} \cdot n_B + \sum_{i=1}^{C} h_{B,DD} \quad [bytes], 
\end{equation}
\vspace{-5pt}
\begin{equation*}
T_{GD-van.}(C) = {\cal B}_{GD} \cdot k_B + \sum_{i=1}^{C} \big(h_{B,GD-van.} + d_B(i)\big)\quad [bytes], 
\end{equation*}
\vspace{-5pt}
\begin{equation*}
T_{GD-red.}(C) = {\cal B}_{GD} \cdot k_B + \sum_{i=1}^{C} \big(h_{B,GD-red.} + d_B(i)\big)\quad [bytes],
\end{equation*}
\vspace{-5pt}
\begin{equation*}
T_{GD-dual}(C) = {\cal B}_{GD} \cdot n_B + \sum_{i \in Q} d_B(i) + \sum_{i=1}^{C} h_{B,GD-dual}\quad [bytes],
\end{equation*}

\noindent
and $Q$ is the set of chunks for which there is no exact chunk at the sink, but for which there is a basis, \ie where GD could find a match that is not an exact one.
The total number of elements in $Q$ is $B_{DD}-B_{GD}$.
We consider $\sum_{i=1}^{C} \big(h_{B,GD-red.} + d_B(i)\big) = \sum_{i=1}^{C}h_{B,DD}$ and $h_{B,DD} = h_{B,GD-van.} = h_{B,GD-dual}$, where $h_{B,GD-dual}$ is the addition length of $f_{b_i}^\prime$ and $f_{c_i}^\prime$.
Considering $C \cdot n_B$ as the original size of data, we calculate the compression ratio as $\frac{C \cdot n_B}{T_{\cal P}}$.

Notice that
\begin{equation*}
T_{DD}(C) < C \cdot n_B \Rightarrow {\cal B}_{DD} < \frac{n_B - h_{B,DD}}{n_B} \cdot C.
\end{equation*}

We can now calculate the conditions for which GD will outperform DD for the different schemes as follows.

\textbf{GD-vanilla:}
For GD-vanilla to outperform DD, we assume:
\begin{equation}
\begin{split}
&T_{GD-van.}(C) < T_{DD}(C) \Rightarrow\\
&{\cal B}_{GD} <  \frac{{\cal B}_{DD} \cdot n_B - \sum_{i=1}^{C}d_B(i)}{k_B}.
\end{split}
\label{eq:GDoutDD}
\end{equation}

We consider $M_{GD} = C - {\cal B}_{GD}$ and ${\cal M}_{DD} = C - {\cal B}_{DD}$ as the total number of matches for GD and DD, respectively, out of $C$ data chunks.
Using \autoref{eq:GDoutDD}, we can determine that GD-vanilla outperforms DD if:

\begin{equation*}
\begin{split}
&(C - M_{GD}) \cdot k_B + \sum_{i=1}^{C}d_B(i) < (C - M_{DD}) \cdot n_B \Rightarrow\\
&\frac{\sum_{i=1}^{C}d_B(i) + C \cdot (k_B - n_B) + M_{DD} \cdot n_B}{k_B} < M_{GD}
\end{split}
\end{equation*}

For the cases where $\sum_{i=1}^{C} d_B(i) + C \cdot k_B = C \cdot n_B$, GD-vanilla outperforms DD if the number of matches for GD is greater than $\frac{n_B}{k_B}$ times the number of matches for DD.
Notice that the number of matches for GD is always equal or greater than the number of matches for DD. 

\textbf{GD-reduced:}
Using a similar analysis, we can determine that GD-reduced always improves the transmission cost compared to DD, because $k_B<n_B$ and $\sum_{i=1}^{C} \big(h_{B,GD-red.} + d_B(i)\big) = \sum_{i=1}^{C}h_{B,DD}$.

\textbf{GD-dual:}
By replacing $B_{DD}$ with $B_{GD}+(B_{DD}-B_{GD})$ in \autoref{eq:DDcost} and considering that for any $i$, $d_B(i)<n_B$, shows that GD-dual reduces the transmission cost compared to $DD$ in all scenarios.

\emph{Example.}
For $n_B=128$, $n_k=127$, $h_B=20$ and $C = 1000000$, DD reduces the transmission cost if $M_{DD}>156250$.
If $\sum_{i=1}^{C} d_B(i) + C \cdot k_B = C \cdot n_B$, GD-vanilla outperforms DD as long as $M_{GD} > 1.00787M_{DD}$ which means GD should have at least $1231$ more matches.
However if $\frac{\sum_{i=1}^{C} d_B(i)}{C} + k_B - n_B = 1$ then GD-vanilla outperforms DD if $M_{GD} > 1.00787 M_{DD} + 7874.01$ which means GD should have at least \num{9105} more matches compared to DD.

\subsection{Compression ratio}
\label{sec:compressionrate}

As shown in \autoref{fig:compression}, we achieve different compression ratio for LZW, DEFLATE~\cite{RFC1951-DEFLATE}, DD and the various GD variants over a real-world data for different chunk lengths. %

All schemes are applied over the data at byte level. %
We consider 6-byte fingerprints ($h_{B,DD} = 6$) and Hamming codes as the transformation.

We observe that GD-reduced and GD-dual (with or without offset removal) outperform LZW, DEFLATE and DD for chunk lengths of $16$, $64$ and $128$ bytes.
For large chunk length of $256$~B, DEFLATE provides better compression ratio compared to other schemes while our techniques outperform LZW and DD with gains of up to $1.41\times$ and $1.48\times$, respectively.
GD-vanilla with offset removal provides up to $1.4\times$ and $1.05\times$ better compression ratio compared to DD for chunk lengths of $256$ and $128$~B, respectively.

\subsection{Transformation function effect}

\autoref{tab:basisCounter} compares the number of unique basis under DD, GD using Hamming with and without offset removal, and using Reed-Solomon as transformation functions considering $2$ real-wold data sets.
In addition to the water dataset~\cite{water}, we use energy measurements from private households~\cite{ECO,beckel2014eco} using readings spanning a 6-month time span.\footnote{From 01.07.2012 until 31.01.2013.}

We applied DD and GD over the columns on byte level.
\autoref{tab:basisCounter} shows that the number of unique bases for GD is lower than the one for DD for all the schemes which means that the number of matches for GD is greater than the number of matches for DD for both datasets. 
GD based on Hamming with offset removal has reduced the number of unique bases by up to $35\%$ for water meter measurements. 

\begin{table*}[t!]
   \centering
   \begin{threeparttable}[b]
   \caption{Number of bases for real-world data sets.}
   \rowcolors{1}{gray!10}{gray!0}
   \label{tab:basisCounter}
   \begin{tabular}{r|r|r|r|r|r|r|r|r|r|r}

   \rowcolor{gray!60}
   \multirow{3}{*}{} & \multicolumn{5}{c|}{\textbf{Water meter measurements}} & \multicolumn{5}{c}{\textbf{Electricity consumption} \& \textbf{occupancy}}\\

   \rowcolor{gray!40}
  & \multirow{2}{*}{} & \multicolumn{4}{c|}{\# Bases} & \multirow{2}{*}{} & \multicolumn{4}{c}{\# Bases}\\

   \rowcolor{white}
   $n_B$ & \# Chunks & DD & \makecell{GD \\ Hamming} & \makecell{GD\\ Hamming\\offset rem.\tnote{1}} & \makecell{GD \\ $RS$\tnote{2}} & \# Chunks & DD & \makecell{GD \\ Hamming}  &  \makecell{GD \\Hamming \\offset rem.\tnote{1}} & \makecell{GD \\ $RS$\tnote{2}} \\
	\hline
	32 & 8,062,520 & 535,959 & 516,455 & 488,738 & 485,714 & 20,909,713 & 9,971,873 & 9,933,732 & 9,407,828 & 9,898,888\\

	64 & 4,031,260 & 539,377 & 537,689 & 499,233 & 537,010 & 10,454,862 & 5,437,915 & 5,428,332 & 5,246,798 & 5,425,244 \\

	128 & 2,015,630 & 335,678 & 334,244 & 269,619 & 334,877 & 5,227,436 & 2,920,964  &  2,916,975 & 2,852,228 & 2,918,966 \\

	256 & 1,007,815 & 260,276 & 259,720 & 168,140 & 260,534\tnote{3} & 2,613,721 & 1,560,935 & 1,558,771 & 1,532,526 & 1,557,294\tnote{3} \\

	512 & 503,908 & 153,245 & 153,118 & 121,073 & - & 1,306,867 & 832,608 & 831,308 & 821,580 & - \\

	1024 & 251,954 & 90,636 & 90,586 & 67,188 & - & 653,533 & 446,218 & 445,060 & 442,070 & - \\
	\hline
   \end{tabular}
   $^1$ With preprocessing step of \textit{offset removal}. \quad
   $^2$ $RS_{256}(n_B,n_B-2)$. \quad
   $^3$ $RS_{256}(255,253)$.

  \end{threeparttable}
\end{table*}

\section{\label{sec:syst-design-arch}\sys Architecture}
{\label{sec:hermes}}

\emph{Assumptions}. In order to design and deploy a distributed network using \sys, we assume the following:
\emph{(1)} all nodes use the same fingerprint length;
\emph{(2)} all generalized deduplication nodes use the same transformation configuration; and 
\emph{(3)} all deduplication node use the same chunk length, to avoid the need for transmitting the fingerprint/chunk length and the transformation configuration in the network.

\newcommand{\YES}{\textcolor{OliveGreen}{\ding{51}}}
\newcommand{\NO}{\color{BrickRed}{\ding{55}}}
\begin{table}[!t]
\renewcommand{\arraystretch}{1.2}
  \centering
  \small
  \caption{\label{tab:hermes_msg_t}Message types for the \sys protocol and relation with the node classes.}
  \rowcolors{1}{gray!10}{gray!0}
  \begin{tabular}{lccc}
    \hline
    \textbf{Message Type} & \textbf{basic} & \textbf{dedup.} & \textbf{gen. dedup.} \\ \hline
    Response                & \YES & \YES & \YES \\
    Data                    & \YES & \faExternalLink  & \faExternalLink \\
    Deduplication             & \NO  & \YES & \NO \\
    Deduplication data          & \NO  & \YES & \NO \\ 
    Gen. deduplication            & \NO  & \NO    & \YES\\ 
    Gen. deduplication data         & \NO  & \NO    & \YES\\
    \hline
  \end{tabular}
\end{table}

A distributed deployment of \sys allows for different types of nodes: \emph{source}, \emph{sink} and \emph{intermediate} nodes.

Source nodes inject data into the network, while sink nodes only ingest data without further retransmissions.
An intermediate node is any node between a sink and source.

Each node handles messages according to its own (unique) class: \emph{basic}, \emph{deduplication} and \emph{generalized deduplication}.

Basic nodes serve as a pass-through, without any data processing.
Deduplication nodes perform DD on the node.
Finally, a generalized deduplication node performs GD locally.
Depending on its type, a node handles different kinds of messages, as shown in \autoref{tab:hermes_msg_t}.
All nodes can send response messages, used to communicate success, acknowledgement or failures in the system to the previous node in the communication chain.
Basic nodes acting as source can only send raw data using the data message type.

\begin{figure}[!t]
  \centering
    \begin{sequencediagram}
      \newthread{A}{Source}
      \newinst[1]{B}{Intermediate}
      \newinst[1]{C}{Sink}      
      \begin{call}
        {A}{$m_{d}$}{B}{$m_{ack}$}
      \end{call}
      \begin{call}
        {B}{$m_{gd}$}{C}{$m_{nf}$}
      \end{call}
            \begin{call}
        {B}{$m_{gdd}$}{C}{$m_{ack}$}
      \end{call}
    \end{sequencediagram}
  \caption{\label{fig:com-basic-gd-gd}Data transmission from a \texttt{basic} source through a \texttt{generalised deduplication} intermediate to a \texttt{generalised deduplication} sink.}
\end{figure}

A deduplication node can send two types of messages, \emph{(1)} a deduplication message with the chunk finger, and \emph{(2)} a deduplication data message, which piggyback the fingerprint and the chunk itself, when the latter is missing in the system.

A similar pattern is used for generalised deduplication nodes.
Initially, a first deduplication message is send, with the fingerprint of a basis and the deviation.

If the basis is missing in the system, the node sends a deduplication data message, including the fingerprint and the chunk.
If a deduplication or generalized deduplication node receives a data message, they process the message payload using DD or GD, according to their node's type.

To invoke the transmission of (generalised) deduplication data message, a node receiving the message must respond with a \emph{new fingerprint} message as this tells the sending node that it is a basis fingerprint that has not been seen before.

\autoref{fig:com-basic-gd-gd} illustrates the transmission of messages between \sys nodes with a simple 3-nodes topology.
The intermediate and sink nodes use generalized deduplication.

The source sends a data message ($m_d$) to the intermediate node.
Upon reception, the intermediate replies with an \emph{acknowledgement} ($m_{ack}$).

The intermediate node, using GD on the message payload, will construct a generalized deduplication message ($m_{gd}$) and send it to the sink node.
The sink node detects a new fingerprint and responds with the corresponding response message ($m_{nf}$).
In turn, this triggers the intermediate node to send a generalized deduplication data message ($m_{gdd}$), to which the sink will respond with an $m_{ack}$. 

Apart from the computational resources needed to operate GD or DD, nodes must maintain locally a record of seen fingerprints.
Here we keep a continuously growing record of fingerprints in memory.
The memory cost for a single chunk/basis record is $|fingerprint|$ in bytes plus 8 bytes used for data referencing; we call this cost $m$.  
From this, the total number of chunks/bases we can represent in $n$ MB of memory can be computed as $\frac{n \cdot 2^{20}}{m}$.
Using CRC-32 hashes as fingerprints this gives us that in 256MB of memory, we can represent 22,369,621 fingerprints.

\textbf{Security considerations.}

An attacker can inject (generalised) deduplication data messages where the fingerprint $f_i$ is not correct for the basis or chunk in the message, potentially corrupting the data in the system.
If such as message is the first for the given $f_i$ the data in system will be corrupt.
A possible mitigation strategy is to use validation checks, where a fingerprint $f'_i$ is generated for the chunk or basis and the validation will be: \emph{if $f_i = f_i'$ then $f_i$ is valid else $f_i$ is invalid.}
If the result is invalid the message should be rejected and the sender of the message could be excluded from the network. 
Notice however that this attack can be launched also against systems using classical deduplication techniques. %

\textbf{Implementation details.}
The current implementation for \emph{Hermes protocol} supports GD-vanilla and GD-reduced.
We are considering GD-vanilla throughout this paper unless otherwise stated. %
The \sys protocol implementation requires support for C++ 2017, is therefore linked against GNU C++ library 6.0.25, and consists of \num{2394} LOC.

\begin{figure*}[t]
  \centering
  \begin{subfigure}[t]{0.33\textwidth}
    \includegraphics[width=\linewidth,trim={1.3cm 6.4cm 2.2cm 7.2cm},clip]{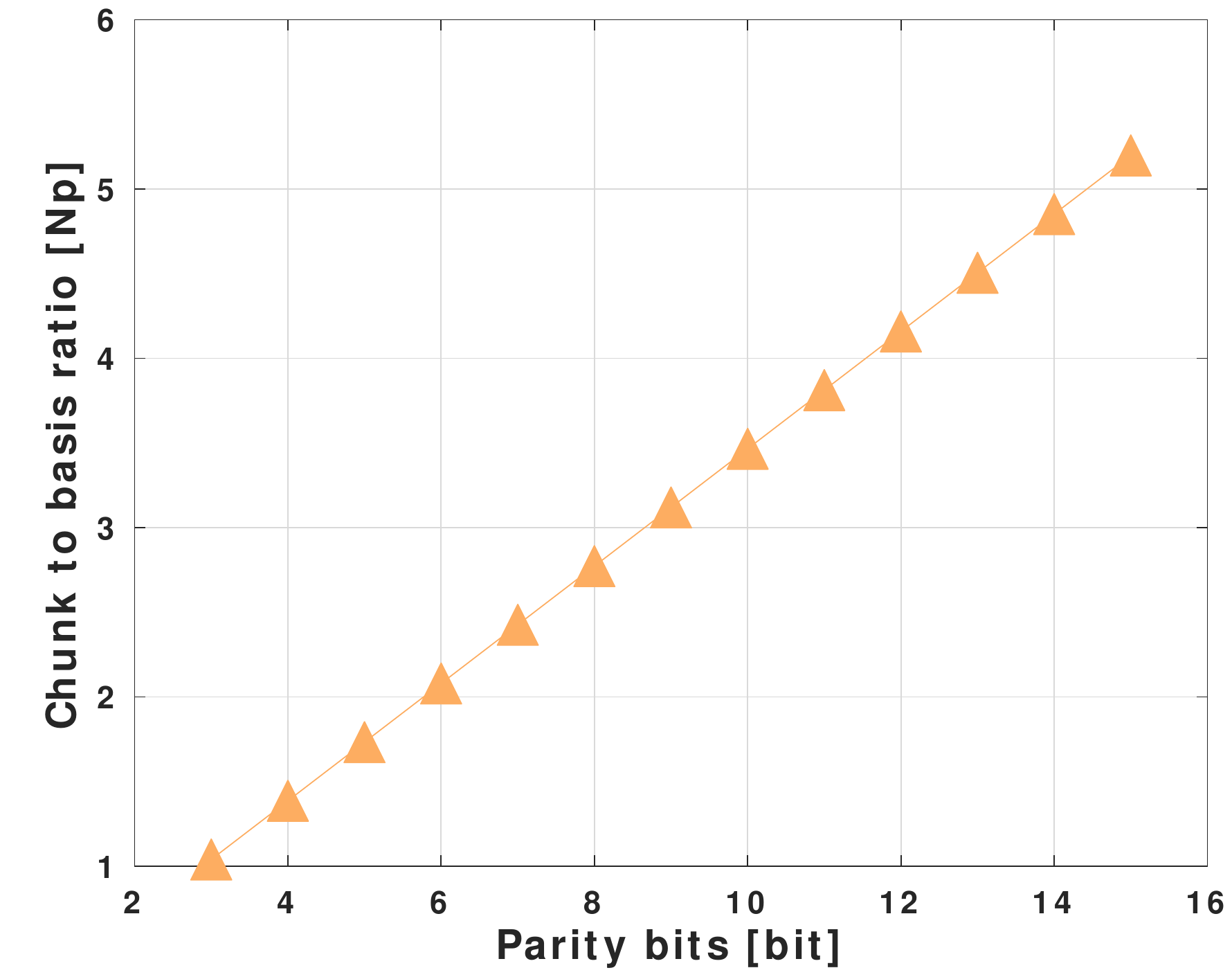}%
    \caption{Chunk to basis ratio\label{fig:c2b}}
  \end{subfigure}%
  \begin{subfigure}[t]{0.33\textwidth}
    \includegraphics[width=\linewidth,trim={1.3cm 6.4cm 2.2cm 7.2cm},clip]{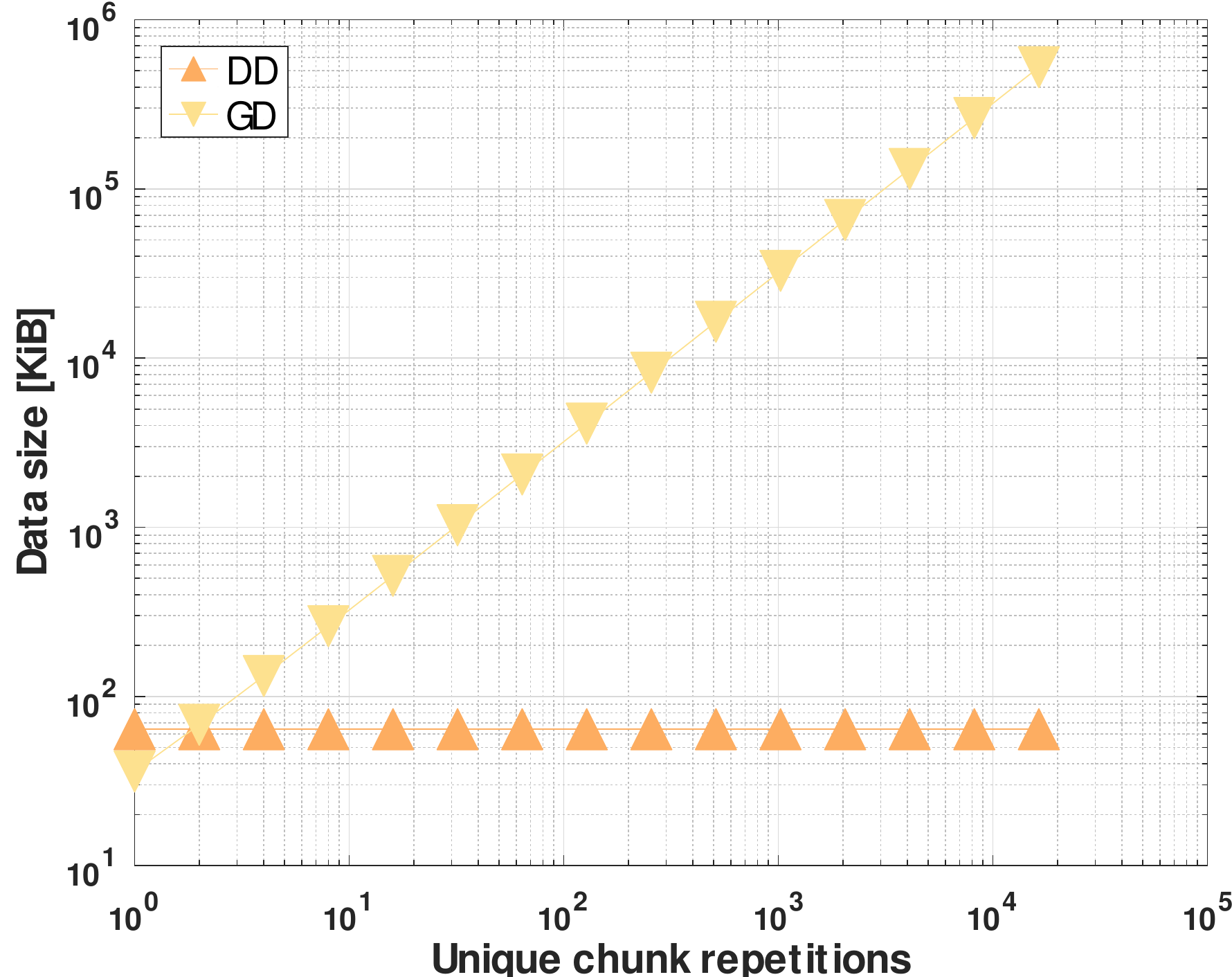}
    \caption{Repetitive chunks\label{fig:repetitions}}
  \end{subfigure}%
  \begin{subfigure}[t]{0.33\textwidth}
    \includegraphics[width=\linewidth,trim={1.3cm 6.4cm 2.2cm 7.2cm},clip]{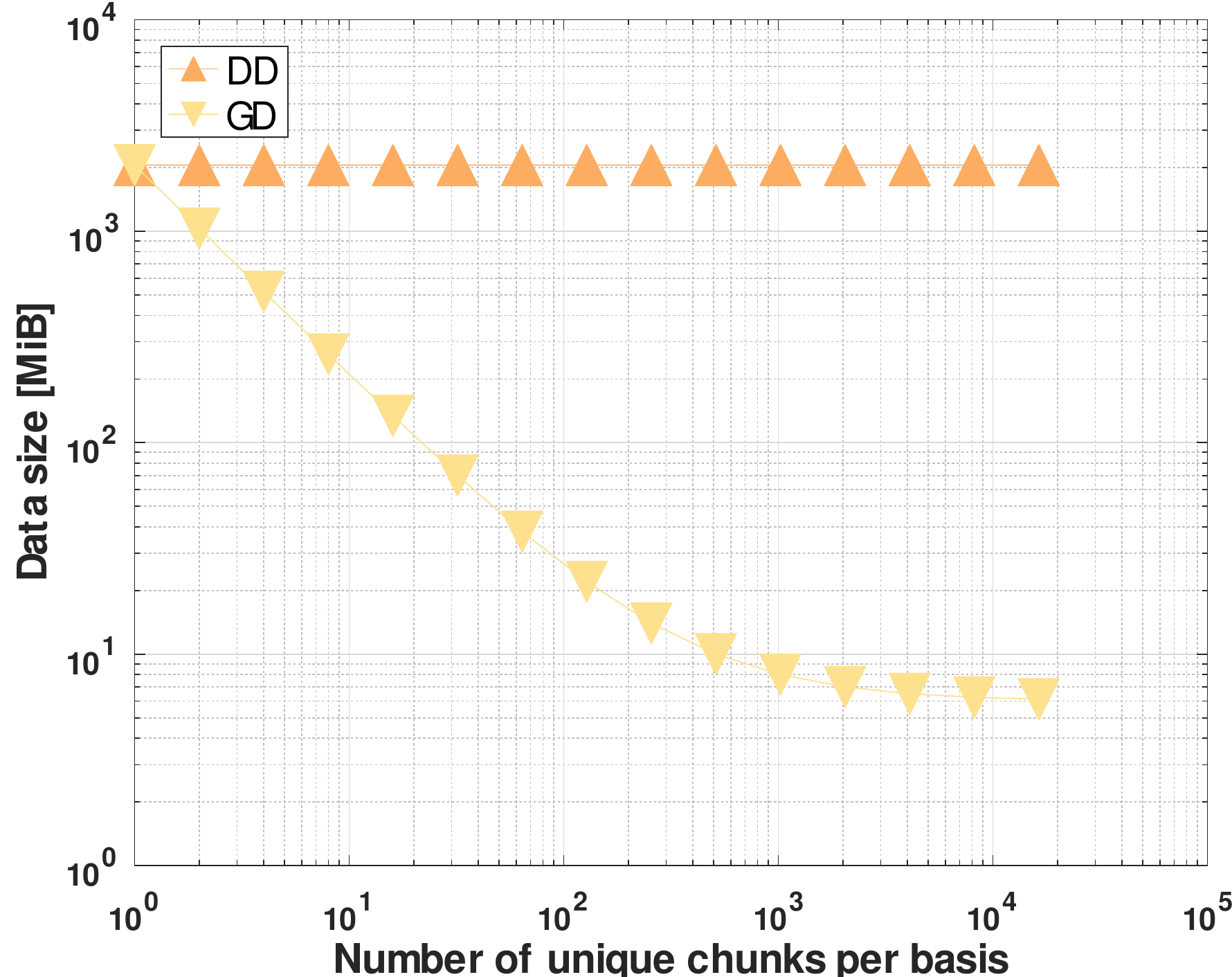}
    \caption{Unique chunks per basis\label{fig:ucpb}}
  \end{subfigure}
  \caption{Comparison of DD and GD.\label{fig:DDvsGD}}
\end{figure*}

\section{Evaluation}{\label{sec:evaluation}}
This section presents our experimental evaluation of the \sys prototype.
Given the intricacies of obtaining accurate power measurements on IoT devices, we begin by describing our experimental and measurement setup.
Finally we present our results based on micro-benchmarks and macro-benchmarks on synthetic datasets.

\subsection{Testbed}
Our experiments are deployed over a switched cluster of 16 Raspberry Pi 4B\footnote{\url{https://www.raspberrypi.org/products/raspberry-pi-4-model-b/}} featuring a Raspberry Pi PoE-HAT\footnote{\url{https://www.raspberrypi.org/products/poe-hat/}} to enable 802.3af Power-over-Ethernet~\cite{mendelson2004all}.

We deploy a simple network topology where Raspberry Pis are used as source nodes and are connected to a Dell PowerEdge R330 server acting as sink node.
The Raspberry Pis are powered using PoE by an Ubiquiti Networks UniFi USW-48P-750 switch and are connected over a Gigabit Ethernet link.

Each Raspberry Pi is running the beta bootloader 2019-12-03 and Raspian Buster Lite (2020-02-13) with a Linux (v4.19.97) which are installed on a \SI{32}{\gibi\byte} SanDisk Extreme microSDXC UHS-I card.

The clocks of all machines used during the benchmarks are synchronized using NTP in order to relate the power consumption to the statistics of a benchmark run.

\subsection{Power Measurements}
The power measurements are gathered using two techniques.
\textbf{PowerSpy2.} The Alciom PowerSpy2\footnote{\url{https://www.alciom.com/en/our-trades/products/powerspy2/}} is a power analyzer connected between a power plug and a power adapter.
It supports two modes: \emph{(1)} data logging, and \emph{(2)} real-time.

The former produces periodic measurements that are stored in the internal persistent memory and can later be fetched and analyzed.
The latter allows the device to stream periodic measurements over Bluetooth v2.

We configure the PowerSpy in real-time mode to record measurements at a frequency of \SI{50}{\Hz} (\ie one new sample is produced every 0.02~s.

\textbf{UniFi Switch.} %

The UniFi switch provides an access protected API reachable over a local \texttt{telnet} connection that can be used to query the PoE status of its ports.
We exploit this option to periodically gather PoE measurements on the utilized ports using an ad-hoc \texttt{expect} script\footnote{\url{https://core.tcl-lang.org/expect/index}}.

With this method our script is able to record PoE information at a frequency of about \SI{8}{\Hz}.
We point out that the switch can sometimes detect the PoE state wrongly, \ie a device is connected but detected by the switch as an open circuit or unknown state.
In these situations the switch does not provide any PoE status information and the power consumption is not measured.

\textbf{Reduction of interferences.} During the execution of the micro-benchmark the Raspberry Pi is attached to the PowerSpy.
With an auxiliary machine we connect simultaneously to the Raspberry Pi for monitoring and to the PowerSpy for recording power measurements.
We interface with the Raspberry Pi via a serial channel (USB-to-UART). %
We employ UART in order to keep the static power consumption of the Raspberry Pi as low as possible and to avoid power interference from other peripherals.
Hence, no peripherals were attached to the Raspberry Pi except for Ethernet and the UART GPIO pins.

Additionally, several system services (\eg  \texttt{cron}, \texttt{ssh},
\texttt{timesyncd}, \etc) are disabled for the micro-benchmark to avoid interference from other processes during the measurement.

\begin{figure*}[t]
  \centering
  \begin{subfigure}[t]{0.33\textwidth}
    \includegraphics[width=\linewidth,trim={1.3cm 6.4cm 2.2cm 7.2cm},clip]{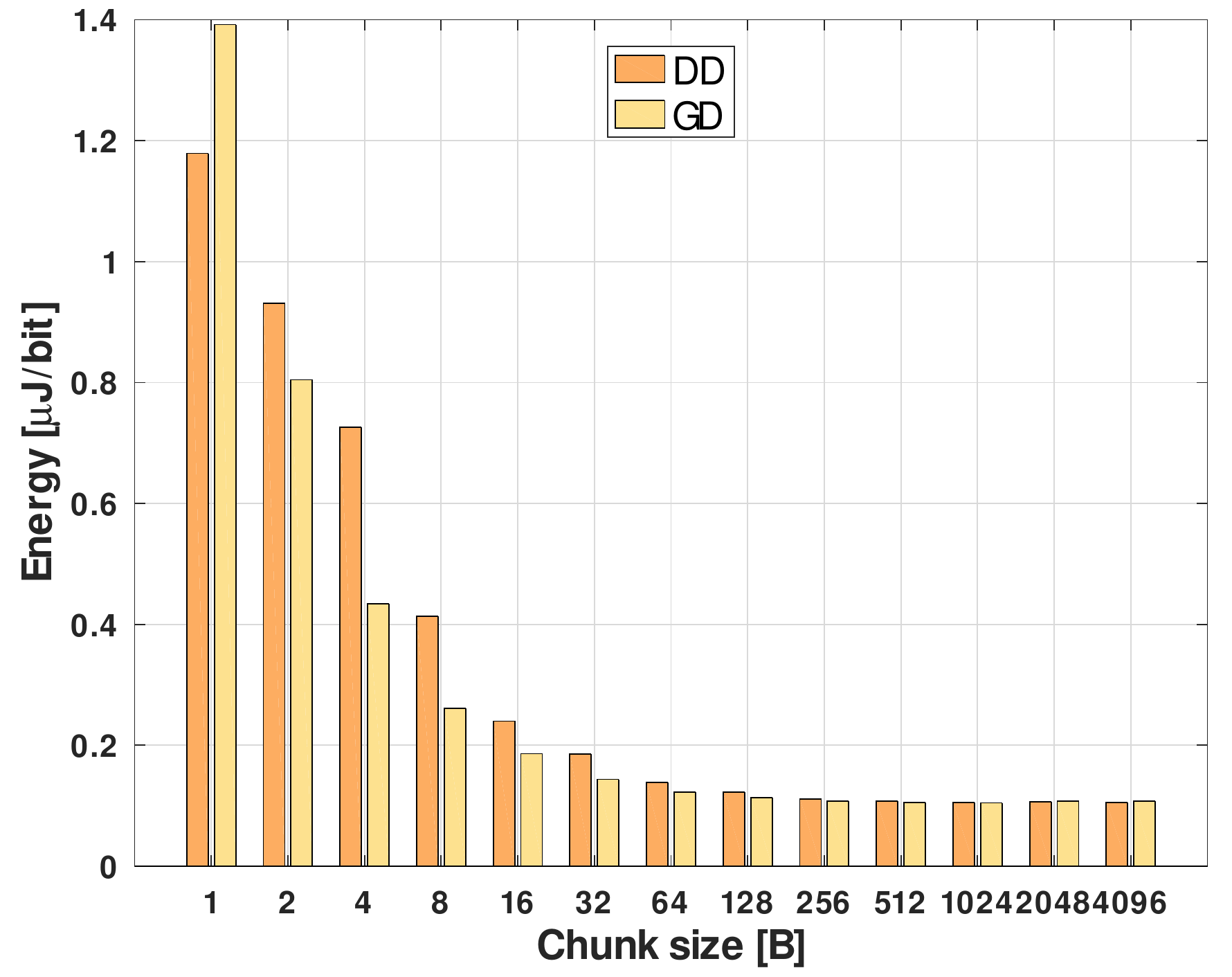}%
    \caption{Energy per bit\label{fig:micro:energy}}
  \end{subfigure}%
  \begin{subfigure}[t]{0.33\textwidth}
    \includegraphics[width=\linewidth,trim={1.3cm 6.4cm 2.2cm 7.2cm},clip]{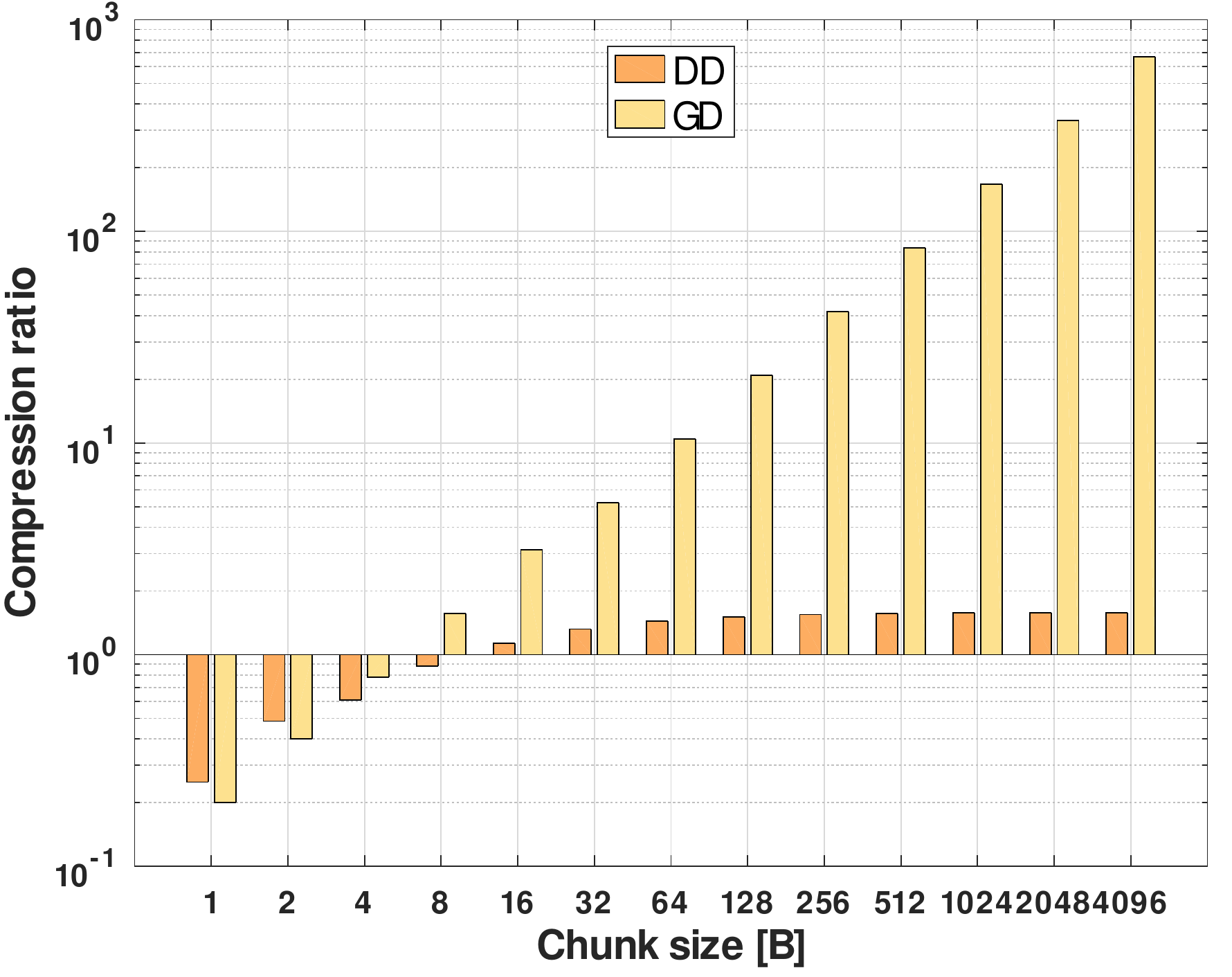}%
    \caption{Compression ratio\label{fig:micro:compress}}
  \end{subfigure}%
  \begin{subfigure}[t]{0.33\textwidth}
    \includegraphics[width=\linewidth,trim={1.3cm 6.4cm 2.2cm 7.2cm},clip]{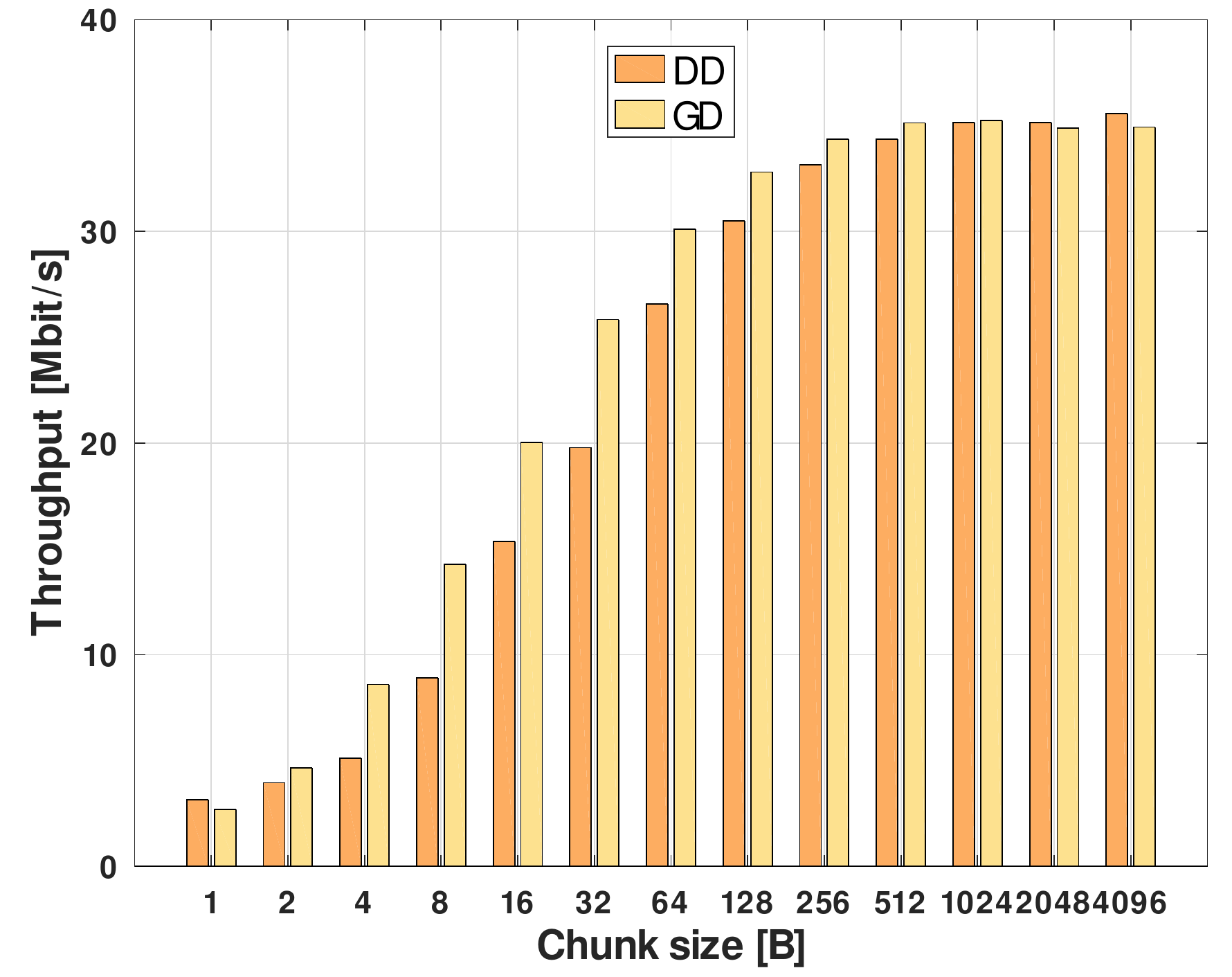}%
    \caption{Throughput\label{fig:micro:tput}}
  \end{subfigure}%
  \caption{Micro-benchmark on Raspberry Pi 4B. \label{fig:micro:rapi4}}
\end{figure*}

\subsection{Classic vs. Generalized Deduplication}
First we demonstrate the impact a data set with certain properties can have on classic and generalized deduplication.
More specifically we analyze the impact of repetitive chunks and the number of chunks mapping to a basis.

We will show our best case scenario using synthetic data sets.
Let us consider Hamming codes as the transformation function.
Then, all chunks which are $1$ bit away from an error-free codeword (our basis) will be mapped to the same basis.
Given this fact, we generate data sets for each chunk length of $n_B = \lceil \frac{2^m}{8} \rceil$ individually, as follows.
First, we generate random error-free codewords.
Then, for each error-free codeword, we generate chunks which are $1$ bit away from it. 

Our generator for the synthetic data set can be parameterized to create a specific number of (unique) bases and to derive a specific number of (unique) chunks.
A basis can be easily generated by selecting a random number of length $k$~bits.
The error-free codeword is made by encoding the basis.
Chunks can then be derived by flipping a random bit and testing that the newly created chunk maps to the same basis.

With generalized deduplication data chunks are mapped to basis.
The ratio of chunks to basis for consecutive numbers of parity bits is depicted in~\autoref{fig:c2b} and given in nepes.
As the number of parity bits increases (and therefore also the chunk size), so do the number of chunks mapping to a basis.
Consequently the compression ratio increases with the number of parity bits.

\begin{figure}[!t]
  \centering
  \includegraphics[width=0.65\linewidth]{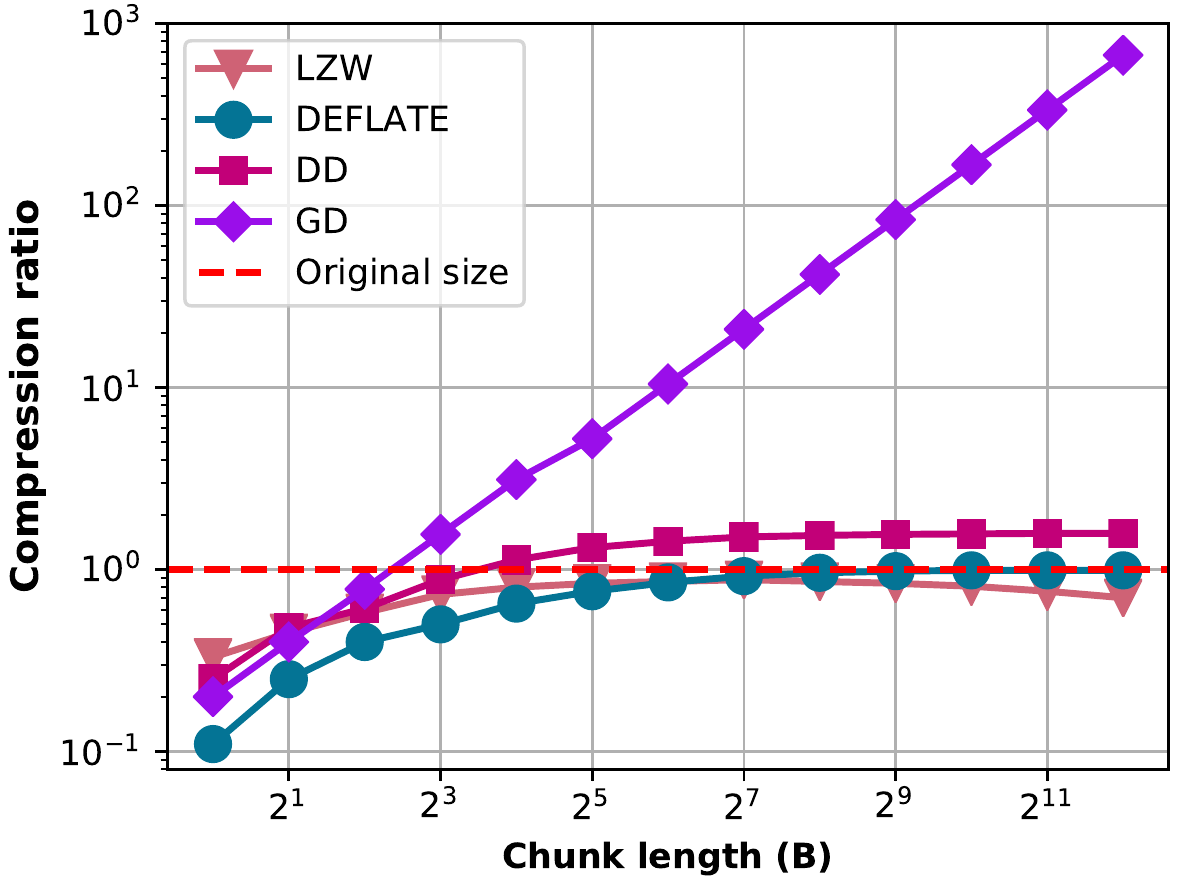}
  \caption{Compression ratio under different schemes using synthetic data sets.}
  \label{fig:best}
\end{figure}

\autoref{fig:best} compares the compression ratio of our technique based on GD with two standard compression algorithms, DEFLATE and LZW, considering the synthetic data sets.
We consider \emph{CRC32} to generate the fingerprints.
We apply all the schemes on each chunk individually.
\autoref{fig:best} shows neither DEFLATE nor LZW compress the data (compression ratio is lower than $1$).
For larger chunk lengths, DD slightly compress the data.
That is while GD compresses the data significantly after the chunk length of 8~B. 
GD provides a compression ratio of $334$ and  $668$ for chunk lengths of $2048$ and $4096$ bytes, respectively. 
Compression ratio for DD is only $1.58$  for chunk lengths of $2048$ and $4096$ bytes. 

Data sets with a lot of repetitive chunks are favored by DD over GD as can be seen for multiple repetitions of the full set of \SI{2}{\byte} chunks in~\autoref{fig:repetitions}.
Although GD is capable of mapping multiple chunks to a basis compared to DD, GD also has to store the deviation of a chunk to its basis.
The information held by the deviation results in an additional storage requirement for GD and is thus an unfavored characteristic of a data set.

The last property we mention is the number of unique chunks per basis.
The number of unique chunks mapping to a basis increases with the chunk size as previously shown in~\autoref{fig:c2b}. 
In~\autoref{fig:ucpb} we demonstrate the different storage requirements for DD and GD for a data set of a little more than a million chunks.
By gradually increasing the number of unique chunks, we see that GD can map more chunks to a basis and achieves a better compression ratio.

\subsection{Micro-benchmark}

Our set of micro-benchmarks are shown in \autoref{fig:micro:rapi4}, using the previously described synthetic datasets.
We show the different trade-offs in terms of chunk-size, energy per bit, compression ratio and throughput.
Results show that energy per bit performance of GD is comparable to DD even considering the added transformation computation.
For chunk sizes of 8 bytes and above, GD not only compresses the data, but it does so significantly better than DD.
For chunks of 64 bytes, GD requires an order of magnitude fewer bits than for DD to represent the data.
For chunks of 4096 bytes, GD outperforms DD by three orders of magnitude.
This is due to the large number of chunks matched to each basis.
Finally, the achieved throughput maxes out at 30+ Mbps for a single thread.
We have considered all the operations to measure this throughput.
These operations include reading the data from memory, applying the compression algorithm and looking for fingerprints in memory to check the availability of the fingerprint for DD and GD.
In a real world scenario, a source node only needs to apply the compression algorithm and it is the sink node which looks for the fingerprint.
Thus, we expect a higher throughput in real deployments.
Future work will consider using NEON instruction sets in ARM to speed up processing.

\begin{figure*}[t]
  \centering
  \begin{subfigure}[t]{0.33\textwidth}
    \includegraphics[width=\linewidth,trim={1.3cm 6.4cm 2.2cm 7.2cm},clip]{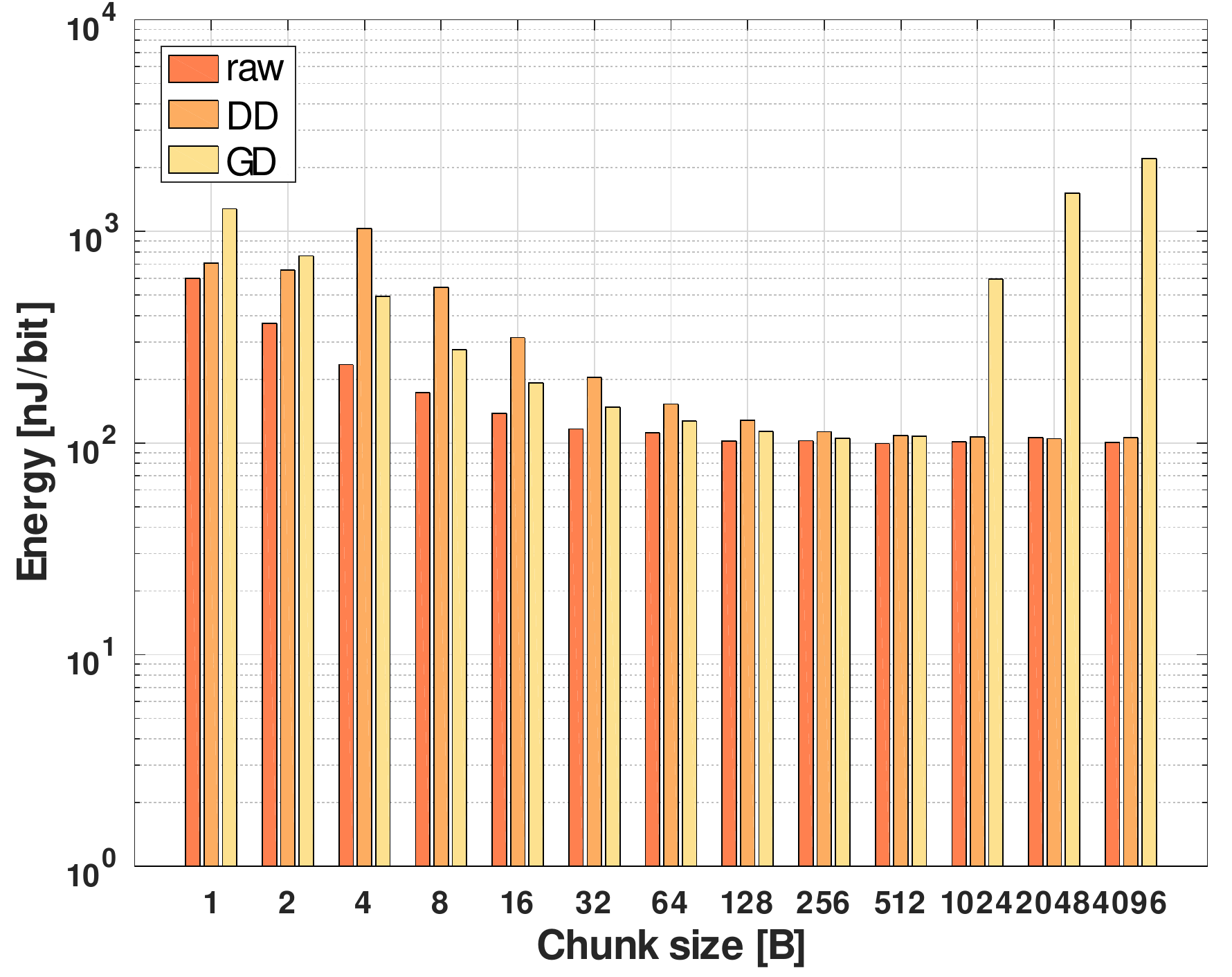}%
    \caption{Energy per bit\label{fig:macro:energy}}
  \end{subfigure}%
  \begin{subfigure}[t]{0.33\textwidth}
    \includegraphics[width=\linewidth,trim={1.3cm 6.4cm 2.2cm 7.2cm},clip]{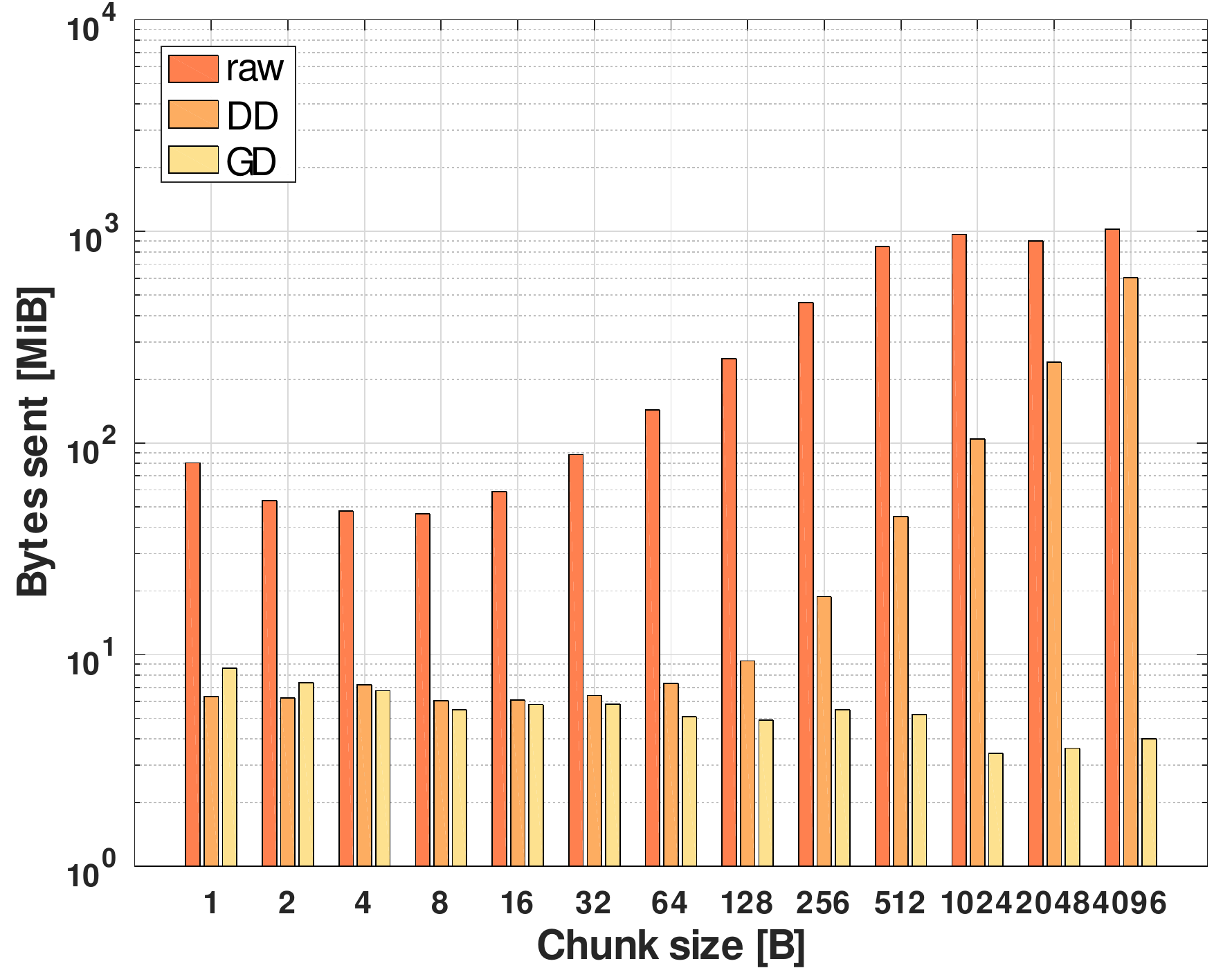}%
    \caption{Network traffic generated\label{fig:macro:sent}}
  \end{subfigure}%
  \begin{subfigure}[t]{0.33\textwidth}
    \includegraphics[width=\linewidth,trim={1.3cm 6.4cm 2.2cm 7.2cm},clip]{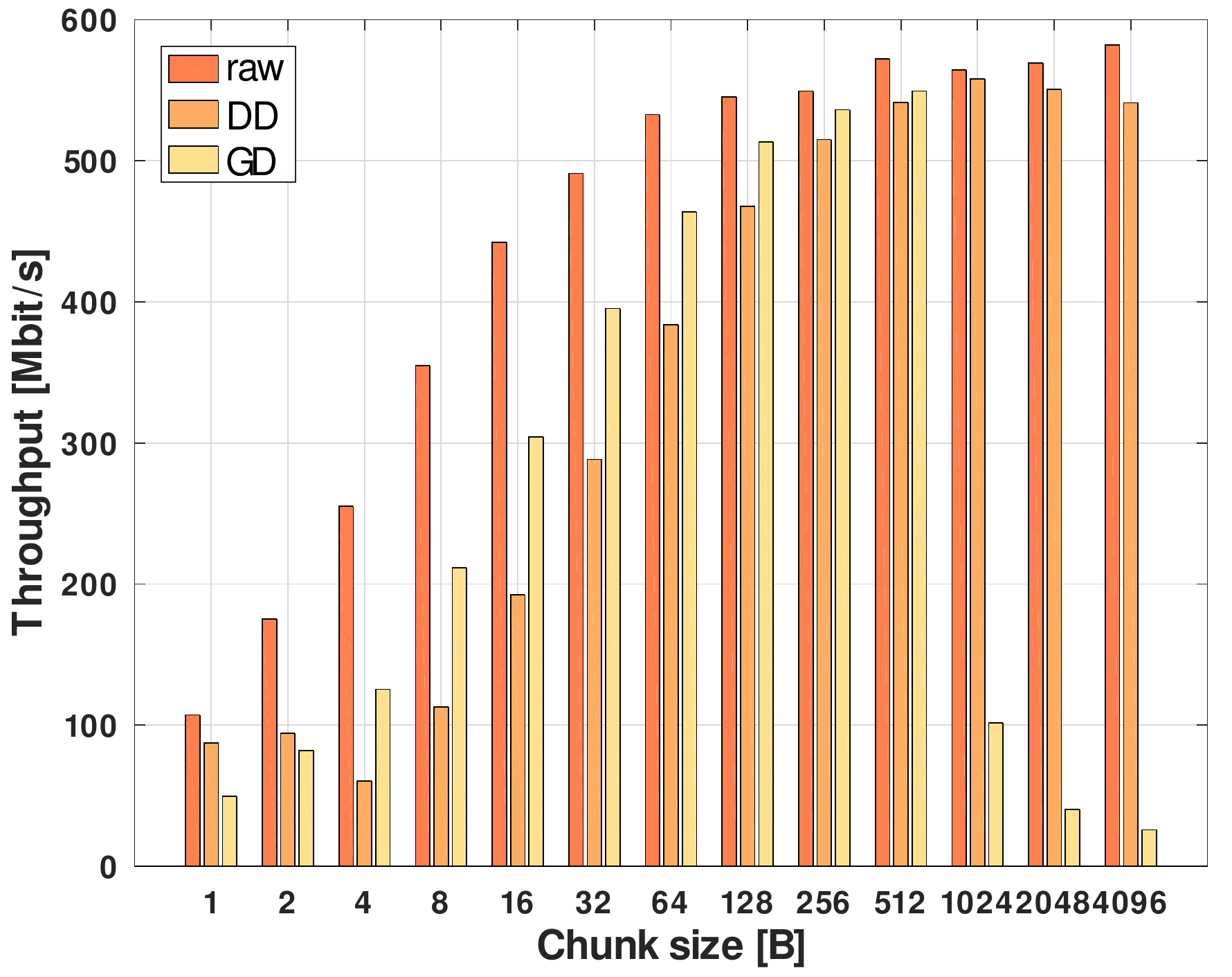}%
    \caption{Throughput\label{fig:macro:tput}}
  \end{subfigure}%
  \caption{Macro-benchmark on Raspberry Pi 4B cluster.\label{fig:macro:rapi4}}
\end{figure*}

\subsection{Macro-benchmark}

In the macro-benchmark, we present the combined statistics for the Raspberry Pi cluster.
We evaluate three different transmission methods: raw, DD and GD.
By raw, we designate sending raw data chunks.
Although our prototype implementation has multi-threading capability, we decided to use only two threads to be comparable to the micro-benchmark.
We use one thread to handle the networking and a second thread to do the necessary transformations and look-ups.
\autoref{fig:macro:energy} shows that for smaller chunks the energy is higher.
This is due to the inaccuracy of the measurement done with the UniFi switch.
The switch has a much lower time resolution to update the PoE status statistics, which we observed to be around 4 to 5 seconds.
With large chunk sizes (larger than typical IoT ones), the energy also increases for GD.
What is most important, is that the network traffic generated for the different methods in \autoref{fig:macro:sent} is reduced significantly with GD.
As seen in the micro-benchmark, the macro-benchmark also shows an optimal throughput between 4 and 12 parity bits.
Approaching 1024 KiB, we recognize a significant drop in throughput for GD.
We assume this is due to hardware limitations, like exceeding cache limits.

\section{Related Work}
\label{sec:relwork}

A large number of lossy and lossless compression techniques have been proposed in the literature to reduce the data transmission by source nodes.
Lossy strategies for compressing across multiple data sources, such as compressive sensing~\cite{Haupt2008,Meng2009,Cheng2010}, error correcting codes to achieve distributed source coding~\cite{Pradhan2003,Eckford2005,elzanaty2019lossy}, usually provide high data compression at the cost of introducing some distortion (losses, errors) at the time of reconstruction of the data.
However, for many IoT applications, such as smart metering or industrial sensing,  errors are not acceptable.
These applications usually rely on a combination of standard compression algorithms applied to a packet-by-packet basis and delta compression within the same packet.
Applying standard compressors to small data limits their compression potential.
Given the limited computing and memory capabilities of most IoT devices, these devices rely on simpler algorithms (\eg, LZW) rather than more advanced but more computationally and memory intensive ones (\eg, DEFLATE~\cite{RFC1951-DEFLATE}, 7z~\cite{7zip}).

Deduplication (DD) is used widely to reduce both storage and transmission cost~\cite{pooranian2018rare}.
Applying DD at the sink node reduces the storage cost by removing equal chunks or files, an approach usually called \emph{target-based}~\cite{harnik2010side}.
Instead, when applied at the source node, \ie, \emph{source-based} approach~\cite{harnik2010side}, DD reduces simultaneously both storage and transmission cost.

\emph{Cross-user} deduplication helps to reduce the storage cost further by deduplicating data over all the users' data.
It has serious security implications due to deterministic status response to the existence of a chunk~\cite{harnik2010side}.
To solve this issue, a wide range of approaches have been proposed~\cite{gao2020secure,zhang2019towards,pooranian2018rare}.
However, all these solution generally suffer from high computational complexity and/or negatively impact the compression rate.

In contrast to DD, \emph{source-based, cross-user} GD does not provide a deterministic response regarding availability of a chunk.
GD expands a chunk into a basis and a deviation to increase the hit probability into previously received bases.
For example, using Hamming as the transformation function for GD and for chunk length of $1$\,KB, there exist $8192$ different chunks mapped to each basis.
The number of possible chunks increases when considering Reed-Solomon as the transformation function.

\section{Conclusion and Future Work}
\label{sec:conclusion}

This work proposed and evaluated a new protocol (\sys) for data compression across multiple sources using the emerging concept of generalized deduplication.
GD generalizes the concept of deduplication by introducing a systematic, transformation stage that allows us to cluster similar data without the need to compare it to a pool of values (as in Delta encoding) or using similarity fingerprints.

We have shown that \sys allows the system to reach significant benefits in compression of the data and reducing data transmission without loss.
We achieve this with a small added computational overhead as shown in deployments with Raspberry Pi model 4B.
This evaluation was carried out considering different data chunk sizes and transformation functions for GD.
For small data packets, our evaluations show that GD significantly outperforms standard compression approaches used in IoT scenarios, \eg, LZW, and even more computationally intensive approaches such as DEFLATE.
Additionally, we demonstrated that \sys under ideal conditions can provide orders of magnitude better compression than DD and even LZW/DEFLATE (on a packet by packet basis) given GD's ability to automatically identify similar data.
Finally, we contributed with new and efficient strategies of GD that advance the theoretical work in~\cite{yazdani2019protocols} to address practical considerations and reduce some overheads in GD compared to DD.

Future work will focus on developing transformations that are more data-aware (\eg, dynamically choosing transformations) to enhance overall compression or system performance.
We will also focus on carrying out large-scale deployments of \sys, potentially considering more computationally limited devices, \eg, Arduinos.
Finally, we will enhance \sys to allow its operation over unreliable transport protocols, \eg, UDP, expanding its capability ranging from packet loss correction and management, even considering a joint design with forward error correction techniques and network coding~\cite{Rec2019}, to congestion control.
 \section*{Acknowledgements}
This work was partially financed by the SCALE-IoT Project (Grant No. 7026-00042B) granted by the Independent Research Fund Denmark, by the Aarhus Universitets Forskningsfond (AUFF) Starting Grant Project AUFF- 2017-FLS-7-1, and Aarhus University's DIGIT Centre.

The research leading to these results has also received funding from the European Union's Horizon 2020 research and innovation programme under the LEGaTO Project (\href{https://legato-project.eu/}{legato-project.eu}), grant agreement No~780681.
\bibliographystyle{ACM-Reference-Format}
\bibliography{references}
\end{document}